\documentclass[%
 reprint,
 amsmath,amssymb,
 aps,
]{revtex4-2}

\usepackage{dcolumn}
\usepackage{bm}
\usepackage{euscript,amssymb}
\usepackage{amsthm}
\usepackage{graphicx}
\usepackage{amsfonts}
\usepackage{amsmath}
\usepackage{amssymb}
\usepackage{fancyhdr}
\usepackage{epsfig}
\usepackage{color}
\usepackage{graphicx}
\usepackage{physics}
\usepackage[usenames,dvipsnames]{xcolor}
\usepackage{amssymb}
\usepackage{amsmath}
\usepackage{lipsum}
\usepackage{wrapfig}
\usepackage[all]{xy}
\usepackage{xcolor}
\usepackage[export]{adjustbox}
\usepackage{mathtools}
\usepackage[normalem]{ulem}

\usepackage{url}
\usepackage{hyperref}
\usepackage{natbib}  

 \definecolor{lblue}{rgb}{0.2,0.5,0.75}
 \definecolor{dgreen}{rgb}{.1,.6,.1}
 \definecolor{aw}{rgb}{0.2,0.5,0.75}
  \definecolor{MyDarkRed}{rgb}{0.71,0.14,0.07}
\definecolor{MyDarkBlue}{rgb}{0.1,0,0.7}
\definecolor{MyDarkGreen}{rgb}{0.21,0.44,0.22}
\definecolor{MyGreen}{rgb}{0.21,0.64,0.18}

\newcommand\jcap{J. Cosmology Astropart. Phys.} 
\newcommand\mnras{MNRAS}
\def\lsim{\mathop{\hbox{${\lower3.8pt\hbox{$<$}}\atop{\raise0.2pt\hbox{$\sim$}}
$}}} \def\gsim{\mathop{\hbox{${\lower3.8pt\hbox{$>$}}\atop{\raise0.2pt\hbox{$
\sim$}}$}}}  

\newcommand{\ud}{\mathrm{d}}

\newcommand{\CR}{{\cal R}}

\usepackage{hyperref}
\hypersetup{
    colorlinks=true,
    citecolor=blue,
    linkcolor=magenta,
    filecolor=cyan,      
    urlcolor=cyan,
}

\begin{document}

\preprint{APS/123-QED}

\title{Relativistic modeling of cosmological structures with Bianchi IX spacetimes}

\author{Przemys{\l}aw Ma{\l}kiewicz}
\email{Przemyslaw.Malkiewicz@ncbj.gov.pl}

\author{Jan J. Ostrowski}
\email{Jan.Jakub.Ostrowski@ncbj.gov.pl}

\author{Ismael Delgado Gaspar}
\email{Ismael.DelgadoGaspar@ncbj.gov.pl}

\affiliation{National Centre for Nuclear Research, Pasteura 7, 02-093
  Warszawa, Poland} 

\date{\today}

\begin{abstract}
We develop a relativistic framework to investigate the evolution of cosmological structures from the initial density perturbations to the highly nonlinear regime. Our approach involves proposing a procedure to match ``best-fit", exact Bianchi IX (BIX) spacetimes to finite regions within the perturbed Friedmann-Lema\^itre-Robertson-Walker universe characterized by a positive averaged spatial curvature. This method enables us to approximately track the nonlinear evolution of the initial perturbation using an exact solution. Unlike standard perturbation theory and exact solutions with a high degree of symmetry (such as spherical symmetry), our approach is applicable to generic initial data, with the only requirement being positive spatial curvature. By employing the BIX symmetries, we can systematically incorporate the approximate effects of shear and curvature into the process of collapse. Our approach addresses the limitations of both standard perturbation theory and highly symmetric exact solutions, providing valuable insights into the nonlinear evolution of cosmological structures.

\end{abstract}

\maketitle


\section{\label{sec:level1}Introduction}
Relativistic modeling of the large scale structures poses serious mathematical difficulties rooted in the nonlinearity of Einstein's field equations. In fact, a limited number of exact solutions with a very often high degree of symmetries forces one to search for approximation methods. Cosmological perturbation theory is one of the most successful examples. However, it suffers from two main drawbacks: (a) As any perturbation theory it operates on the assumption of smallness of a certain parameter (or set of parameters) and thus, by definition, has a limited range of validity. (b) A major, though often overlooked, conundrum with cosmological perturbation theory concerns the notion of background in the general relativity theory or, more specifically, the lack of unambiguous recipe on how such background should be constructed from a generic spacetime. The latter, known as the ``fitting problem" \cite{1987Ellis} laid ground to the idea of averaging Einstein's equations (see e.g. \cite{Buchert1}, \cite{Buchert2} for a relativistic scalar averaging approach).  

As in other theories dealing with vastly complicated systems, one may alternatively perform some form of averaging or coarse graining to reduce the number of degrees of freedom and make the behavior of the system tractable in some approximate sense. The problem of averaging in cosmology has a long history. Two main issues become immediately apparent when the spatial averaging of Einstein's equations is attempted: (a) the notion of spatial integration depends on the chosen foliation of spacetime; (b) there is no useful definition of an average of a tensor. In an ideal situation, one would average an inhomogeneous metric together with the associated stress-energy tensor, and relate them by modified Einstein's equations with terms that account for the backreaction. In the cosmological context, such an averaged metric would necessarily be spatially homogeneous and probably, on the largest scales, would correspond to the Friedmann-Lema\^itre-Robertson-Walker metric. 

While there is no meaningful notion of the average of a tensor, a wide class of spatially homogeneous metrics that admit a three-parameter isometry group are well known and classified into the so-called Bianchi types \cite{Jantzen:2001me} \footnote{It is worth noting that the Bianchi models do not exhaust all the possible homogeneous cosmologies: the Kantowski-Sachs model is a homogeneous cosmology invariant under the group $\mathbf{R}\times SO(3)$ that involves four independent parameters and does not posses any three-dimensional subgroup.}. One could hypothesize that there exist best-fit Bianchi metrics to most of the possible inhomogeneous metrics within a finite region of the Cauchy surface. This idea was
suggested in \cite{1977Spero,1978Spero}, where the authors introduced a method to determine the best-fit three-metric using a variational approach.

In our work we propose a new method to go beyond the regime of validity of cosmological perturbation theory to the nonlinear regime by assigning an exact solution of Einstein's equations to a finite domain within a perturbed FLRW universe. We conjecture that a suitably fitted Bianchi IX (BIX) metric can track the evolution of the initial overdensity. Our choice of the model is guided by the fact that, in the comoving, synchronous coordinate system, a positive spatial curvature is required for the initially expanding domain to change the sign of the scalar expansion and start collapsing, thus making the BIX a natural candidate to emulate this process. Indeed, looking at the Hamiltonian constraint (for a dust model with $G=\frac{1}{16\pi}$),
$$\frac{1}{3}\theta^2 = \frac12 \rho +\sigma^2 -\frac{1}{2}\CR,$$
where $\theta$ is the expansion scalar, $\rho$ is the energy density and $\sigma^2$ is the shear rate, we immediately see that the necessary condition for the turnaround ($\theta=0$) requires a positive spatial curvature $\CR>0$. Our selection of domains to be modeled by the BIX spacetimes is confined to those with a positive averaged scalar curvature (that is not necessarily locally positive), as it was shown with the use of the scalar averaging techniques to be a necessary condition for the collapse \cite{2019Roukema}. Once the domain satisfies our condition, we introduce a triad on it with the structure constants of the BIX metric and express the perturbed, spatial FLRW metric and the second fundamental form in relation to this triad. We calculate the volume-weighted averages of the first and second fundamental forms, and by requiring all nondiagonal components to vanish (which is always achievable on the fixed hypersurface), we determine the corresponding ``best-fit" BIX metric. This approach associates the BIX spacetime with the initially perturbed FLRW within the finite domain, allowing us to approximate the domain's evolution with the fully nonlinear BIX solution. It is worth noting that while BIX has been previously suggested as a collapse model (see, e.g., \cite{Giani}), this work represents the first systematic approach to the problem, involving procedures applicable to generic cosmological initial conditions.

The paper is organized as follows. We review the basic properties of BIX spacetimes in Sec. \ref{sec:BIX}. Section \ref{sec:Ham} is dedicated to the discussion of Hamiltonian cosmological perturbation theory and the gauge-fixing procedure. We perform the ``best-fit" procedure in Sec. \ref{sec:average} to the perturbed dust-filled FLRW initial conditions. Numerical examples are presented in Sec. \ref{sec:examples}. We summarize our results in Sec. \ref{sec:level1}.

\section{\label{sec:BIX}Bianchi IX spacetimes}
The diagonal Bianchi metric expressed in the Misner variables reads \cite{PhysRevLett.22.1071,cmp/1103841345}
\begin{align}\label{b9metric}\begin{split}
\ud s^2=-N^2\ud t^2+e^{2\beta_0}\left(e^{2\beta_++2\sqrt{3}\beta_-}\omega_1^2 \right. \\ \left. +e^{2\beta_+-2\sqrt{3}\beta_-}\omega_2^2+e^{-4\beta_+}\omega_3^2\right),\end{split}
\end{align}
where the lapse function $N$ and the variables $\beta_0$ and $\beta_{\pm}$ are functions of time. The Cartan forms $\omega_a$ satisfy the Maurer-Cartan equation
$$\ud \omega^a=\frac{1}{2}C^a_{~bc}\omega^b\wedge\omega^c,$$
where $C^a_{~bc}$ are the structure constants of a given homogeneity group of the spatial leaves. The Hamiltonian constraint of the Bianchi class A models filled with a perfect fluid (with a barotropic index $w$) takes the following form ($G=\frac{1}{16\pi}$) \cite{Uggla1997DynamicalSI}:
\begin{align}\label{ham}\begin{split}
H_{tot}&=H_g+H_f= \frac{N\mathcal{V}_0e^{-3\beta_0}}{24}\left(-p_0^2+p_+^2+p_-^2\right. \\ &\left. +24e^{4\beta_0}V(\beta_{\pm}) +24e^{3(1-w)\beta_0}p_T\right),\end{split}
\end{align}
where $p_0$, $p_{\pm}$ are the momenta conjugate, respectively, to $\beta_0$, $\beta_{\pm}$ with $\{\beta_0,p_0\}=\mathcal{V}_0^{-1}=\{\beta_{\pm},p_{\pm}\}$ and all the remaining brackets vanishing. We specify the model to be of type IX and set the structure constants,
$$C^a_{~bc}=\frak{n}\epsilon^a_{~bc},$$
where  $\epsilon^a_{~bc}$ is a totally antisymmetric symbol with $\epsilon^1_{~23}=1$, and $\frak{n}>0$ shall be conveniently fixed below. This implies $V(\beta_{\pm}):=V_{IX}(\beta_{\pm})$, where
\begin{eqnarray}
V_{IX}(\beta_{\pm})&=&\frac{\frak{n}^2}{2}e^{4\beta_+}\left(\left[2\cosh(2\sqrt{3}\beta_-)-e^{- 6\beta_+}\right]^2-4\right) \nonumber \\ &:=&W_{IX}(\beta_{\pm})-\frac{3\frak{n}^2}{2},
\end{eqnarray}
where $W_{IX}(0)=0$. The underlying topology of the spatial sections of the BIX is given by a three-sphere with the volume $\mathcal{V}_0=\int_{S^3}\omega_1\wedge\omega_2\wedge\omega_3=\frac{16\pi^2}{\frak{n}^3}$ for $\beta_0=0$. We set the cosmic fluid to be dust with $w=0$. Choosing cosmological time with $N=1$, we obtain the Hamilton equations,
\begin{align}\begin{split}
\dot{\beta}_{0}&=-\frac{e^{-3\beta_0}}{12}p_{0},\\
\dot{p}_{0}&=\frac{e^{-3\beta_0}}{8}\left(-p_0^2+p_+^2+p_-^2-8e^{4\beta_0}V_{IX}\right),\\
\dot{\beta}_{\pm}&=\frac{e^{-3\beta_0}}{12}p_{\pm},~~\dot{p}_{\pm}=-e^{\beta_0}\partial_{\pm}V_{IX}.\end{split}
\end{align}
The dynamics of the isotropic and anisotropic variables are coupled and have to be solved simultaneously. The dynamics toward smaller volumes becomes famously dominated by an oscillatory and chaotic behavior of anisotropy and is described by a sequence of the Bianchi type I solutions joined by the Bianchi type II transitions \cite{doi:10.1080/00018738200101428,reiterer2010bkl}. On the other hand, for large volumes the influence of shear and anisotropy on the dynamics of the Bianchi IX geometry is usually suppressed and the latter has to be investigated on a case-by-case basis. Most importantly, the dust model undergoes a recollapse for any initially expanding configuration.

\subsection{Isotropic collapse}
In the isotropic limit with $\beta_{\pm}=0=p_{\pm}$ and $V_{IX}=-\frac{3\frak{n}^2}{2}$, the closed Friedmann model with the Hamiltonian constraint \eqref{ham} in the more familiar form is retrieved,
\begin{align}\nonumber
H^2=-\frac{1}{6}R_{iso}+\frac{1}{6}\rho_M,
\end{align}
where the Hubble parameter $H=\dot{\beta_0}$, the curvature $R_{iso}=\frac{3\frak{n}^2}{2 e^{2\beta_0}}$, and the dust energy density $\rho_M=p_Te^{-3\beta_0}$. Upon integrating the Friedmann constraint,
\begin{align*}
\frac{\frak{n}}{4}e^{-\beta_0}\ud t=\ud\left(\arcsin\sqrt{\frac{3\frak{n}^2e^{\beta_0}}{2p_T}}\right),
\end{align*}
and introducing the auxiliary parameter $\chi$ such that $\ud\chi=\frac{\frak{n}}{2}e^{-\beta_0}\ud t$, the following parametric solution is obtained:
\begin{align*}
e^{\beta_0}(\chi)=\frac{p_T}{3\frak{n}^2}(1-\cos \chi),~~t(\chi)=\frac{2p_T}{3\frak{n}^3}(\chi-\sin \chi)\;,
\end{align*}
which when expanded in $\chi$ to lowest orders yields the following approximation:
\begin{align}
e^{\beta_0}(t)\simeq\frac{p_T}{6}\left(\frac{9}{p_T}t\right)^{2/3}\left(1+\frac{\frak{n}^2}{20}\left(\frac{9}{p_T}t\right)^{2/3}\right).
\end{align}
This equation describes the dynamics of the closed Friedmann universe as a linear perturbation to the flat Friedmann universe. The latter is obtained for $\frak{n}^2\rightarrow 0$ with $e^{\beta_0}(t)=\frac{p_T}{6}\left(\frac{9}{p_T}t\right)^{2/3}$. The above expansion, now viewed as an expansion in the intrinsic curvature $\propto \frak{n}^2$, can serve to match the perturbed flat Friedmann universe with a homogeneous and positive curvature perturbation to the closed Friedmann universe that undergoes a recollapse. The closed Friedmann model describes the evolution of collapsing masses in Newtonian gravity, where it is known as the top-hat model.

\subsection{Anisotropic collapse}
The complete Hamiltonian constraint \eqref{ham} recast into the Friedmann-like form reads
\begin{eqnarray}
H^2=\frac{1}{3}\sigma^2-\frac{1}{6}R_{ani}-\frac{1}{6}R_{iso}+\frac{1}{6}\rho_M, \nonumber
\end{eqnarray}
where $H=\dot{\beta_0}$ is the Hubble parameter, $\sigma^2=3(\dot{\beta}^2_++\dot{\beta}^2_-)$ is the shear,~~$R_{ani}=-e^{-2\beta_0}W_{IX}(\beta_{\pm})$ is the anisotropic part of the intrinsic curvature,  $R_{iso}=\frac{3\frak{n}^2}{2 e^{2\beta_0}}$ is the isotropic curvature and $\rho_M=p_Te^{-3\beta_0}$ is the dust energy density.

\begin{figure}
    \centering
    \includegraphics[width=0.4\textwidth]{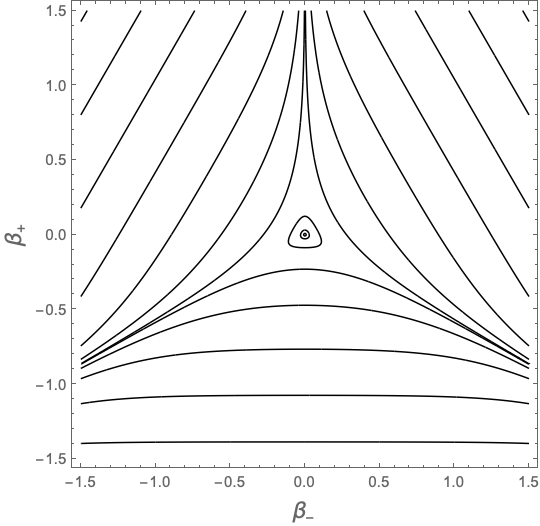}
    \caption{The anisotropy potential $V_{IX}(\beta_{\pm})$ of the Bianchi type IX model.}
\end{figure}
Upon expanding the full Bianchi IX Hamiltonian constraint to first order around the flat Friedmann model, we find:
\begin{gather*}
-2H\delta H-\frac{1}{6}\delta R_{iso}+\frac{1}{6}\delta\rho_M=0,
\end{gather*}
that is, a completely isotropic expression as the linearized shear squared $\sigma\delta\sigma$ vanishes and the linearized intrinsic anisotropic curvature $\delta R_{ani}$ vanishes too. The latter happens because
\begin{eqnarray}
W_{IX}(\beta_{\pm})&=&\frac{\frak{n}^2}{2}e^{4\beta_+}\left(\left[2\cosh(2\sqrt{3}\beta_-)-e^{- 6\beta_+}\right]^2-4\right) \nonumber \\ &+&\frac{3\frak{n}^2}{2} \approx 12\frak{n}^2(\beta_+^2+\beta_-^2)+\dots,
\end{eqnarray}
the anisotropic curvature is at least second order with respect to $\beta_{\pm}$ and first order with respect to $\frak{n}^2$, making it third order. Hence, the anisotropy and shear variables $\beta_{\pm}$ and $p_{\pm}$ do not contribute to the constraint equation. Unlike $\delta R_{iso}$, $\delta H$, and $\delta\rho_M$, they have to be determined from a suitable junction condition (note that $\delta R_{iso}=\frac{3\frak{n}^2}{2 e^{2\beta_0}}$, where $\frak{n}^2$ is a first-order perturbation, is fixed by the free choice of $\frak{n}^2$, whereas $\delta H$ and $\delta\rho_M$ are related by the constraint).

As we rewrite the Bianchi IX three-geometry as a perturbation to the flat Friedmann three-geometry, we find
\begin{eqnarray}\label{pertb9}\begin{split}
\delta \tilde{q}_{11}&=e^{2\beta_0}(2\delta\beta_0+2\delta\beta_++2\sqrt{3}\delta\beta_-)\;, \\
\delta \tilde{q}_{22}&=e^{2\beta_0}(2\delta\beta_0+2\delta\beta_+-2\sqrt{3}\delta\beta_-)\;, \\
\delta \tilde{q}_{33}&=e^{2\beta_0}(2\delta\beta_0-4\delta\beta_+)\;, \\
\delta \tilde{\pi}^{11}&=-\frac{1}{3}e^{-2\beta_0}p_0(\delta\beta_0+\delta\beta_++\sqrt{3}\delta\beta_-)  \\ &+\frac{1}{6}e^{-2\beta_0}(\delta p_0+\frac{1}{2}\delta p_++\frac{\sqrt{3}}{2}\delta p_-)\;, \\
\delta \tilde{\pi}^{22}&=-\frac{1}{3}e^{-2\beta_0}p_0(\delta\beta_0+\delta\beta_+-\sqrt{3}\delta\beta_-)  \\&+\frac{1}{6}e^{-2\beta_0}(\delta p_0+\frac{1}{2}\delta p_+-\frac{\sqrt{3}}{2}\delta p_-)\;,  \\
\delta \tilde{\pi}^{33}&=-\frac{1}{3}e^{-2\beta_0}p_0(\delta\beta_0-2\delta\beta_+) \\ &+\frac{1}{6}e^{-2\beta_0}(\delta p_0-\delta p_+)\;,
\end{split}
\end{eqnarray}
where all the three-metric and three-momentum components are homogeneous. The tildes in $\delta\tilde{q}_{ab}$ and  $\delta\tilde{\pi}^{ab}$ indicate that the respective components are given in the invariant basis of 1-forms and dual vectors. Unlike in the isotropic case, the full set of perturbation variables includes $(\delta\beta_{\pm},\delta p_{\pm})$ beside the isotropic $(\delta\beta_0,\delta p_0)$. They will be inferred from the domain-averaged generic perturbations to the flat Friedmann universe on a suitable hypersurface. To this end, we will express generic metric and expansion perturbations to the flat Friedmann universe in the Bianchi IX-invariant basis.

\section{\label{sec:Ham}Hamiltonian approach to cosmological perturbations}
We expand the Arnowitt-Deser-Misner (ADM) Hamiltonian for gravity and cosmic dust around the flat Friedmann model up to second order,
\begin{align}\label{hamADM}
H_{ADM}\simeq N H^{(0)}+\int_{\Sigma}\ud^3 x\left(N \mathcal{H}^{(2)}+\delta N^\mu\delta\mathcal{H}_\mu\right),
\end{align}
where the gravitational variables are split into the background and perturbation parts: $q_{ij}=a^2\delta_{ij}+\delta q_{ij}$ and $\pi^{ij}=\frac13 p\delta^{ij}+\delta\pi^{ij}$ (with $p=-6\dot{a}$), and analogously the dust variables are replaced with $\phi\rightarrow {\phi}+\delta\phi$ and $\pi_{\phi}\rightarrow\pi_{\phi}+\delta \pi_{\phi}$. $H^{(0)}=-\frac{ap^2}{6}+\pi_{\phi}$ is the background Hamiltonian, with the background variables satisfying $\{a^2,p\}=1=\{\phi,\pi_{\phi}\}$. $N$ is the background lapse function, $\delta N^{0}$ and $\delta N^{i}$ are the perturbations of the lapse function and the shift vector, respectively. The second-order Hamiltonian $\mathcal{H}^{(2)}$ and the linear constraints $\delta\mathcal{H}_\mu$ are functions of the canonical perturbation variables $\delta q_{ab}$, $\delta\pi^{ab}$, $\delta\phi$, and $\delta\pi_{\phi}$, as well as the background quantities $a$, $p$, and $\pi_{\phi}$, and are given in Appendix \ref{adm}. The Hamiltonian $H^{(0)}$ generates the dynamics in the background spacetime, the Hamiltonian $N \mathcal{H}^{(2)}+\delta N_\mu\delta\mathcal{H}^\mu$ generates the dynamics of the perturbation variables, and the linearized constraints $\delta\mathcal{H}^\mu$ constrain the perturbations to physically admissible states.

In the following we switch to the momentum representation for any perturbation variable $\delta X$,
\begin{eqnarray}
\delta\check{X}(\vec{k})=\int_{\Sigma} \delta{X}(\vec{x})e^{-i\vec{x}\vec{k}}d^3x~,
\end{eqnarray} 
where the reality conditions $\delta\check{X}(-\vec{k})=\overline{\delta\check{X}}(\vec{k})$ are assumed. The usual approach splits the perturbations into scalar, vector, and tensor modes by introducing two vectors, $\vec{v}$ and $\vec{w}$, orthogonal to each other and to $\vec{k}$, whose magnitude in the fiducial metric $\delta_{ab}$ reads
\begin{align}
|\vec{v}|=|\vec{w}|=|\vec{k}|^{-1}.
\end{align} 
We assume that such frames are symmetric with respect to the reflection about the origin, that is, when $\vec{k}\rightarrow -\vec{k}$ then $\vec{v}\rightarrow -\vec{v}$ and $\vec{w}\rightarrow -\vec{w}$. We define
\begin{equation}
\delta\check{q}_{ab}=\delta\check{q}_{m}A_{ab}^m,~~~~~~\delta\check{\pi}^{ab}=\delta\check{\pi}^{m}A_{m}^{ab},
\end{equation}
where $A_{ab}^1:=\delta_{ab}$, $A_{ab}^2:=\frac{k_a k_b}{k^2}-\frac{1}{3}\delta_{ab}$, $A_{ab}^3:=\frac{1}{\sqrt{2}}(k_a v_b+k_b v_a)$, $A_{ab}^4:=\frac{1}{\sqrt{2}}(k_a w_b+k_b w_a)$, $A_{ab}^5:=\frac{k^2}{\sqrt{2}}(v_a w_b+v_b w_a)$, and $A_{ab}^6:=\frac{k^2}{\sqrt{2}}(v_a v_b-w_a w_b)$. The matrices $A^{ab}_m$ form the dual basis, i.e. $A^{ab}_mA_{ab}^n=\delta^{n}_m$. The new variables describe, respectively, scalar ($\delta\check{q}_{1}$ and $\delta\check{q}_{2}$), vector ($\delta\check{q}_{3}$ and $\delta\check{q}_{4}$) and tensor ($\delta\check{q}_{5}$ and $\delta\check{q}_{6}$) modes of the metric perturbation. From now on, we confine our analysis to the scalar modes: that is, we put $\delta\check{q}_{j}=0=\delta\check{\pi}_{j}$ for $j=3,4,5,6$. As a result, the Hamiltonian \eqref{hamADM} gets simplified and, in particular, only two out of the four linear constraints remain nontrivial.

\begin{figure}
    \centering
    \includegraphics[width=0.5\textwidth]{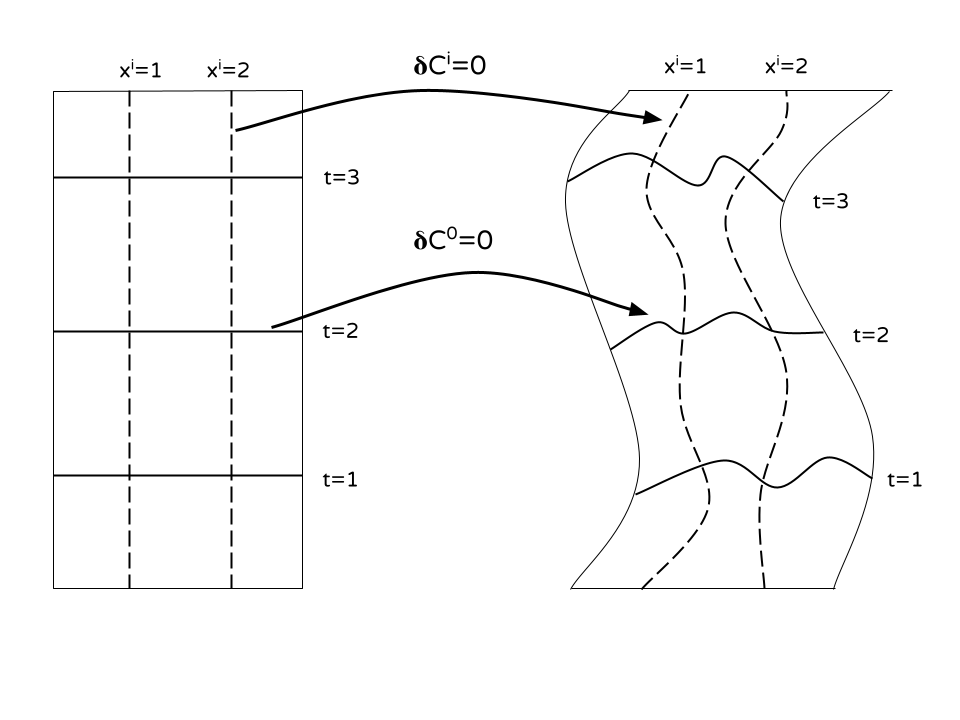}
    \caption{The gauge-fixing conditions $\delta{\mathcal{C}}^0=0$ and $\delta{\mathcal{C}}^{\vec{k}}=0$ fix respectively the time coordinate and the spatial coordinates in the perturbed spacetime model.}\label{coordinates}
\end{figure}

As verified in Appendix \ref{kuch}, the ADM and the dust perturbation variables can be replaced with another set of dynamical variables via a canonical transformation \cite{Boldrin_2022},
\begin{eqnarray}\label{map}\begin{split}
&(\delta\check{q}_{1}, \delta\check{q}_{2}, \delta\check{\pi}_{1}, \delta\check{\pi}_{2}, \delta\check{\phi}, \delta\check{\pi}_{\phi}) \\ 
 &~~~~~~~~~~~~~~~~\downarrow\\
&(\Psi, \Pi_{\Psi}, \delta\mathcal{H}^0, \delta\mathcal{H}^{\vec{k}}, \delta\mathcal{C}^0, \delta\mathcal{C}^{\vec{k}}),\end{split}
\end{eqnarray}
where the new canonical pairs $(\Psi, \Pi_{\Psi})$, $( \delta\mathcal{H}^0, \delta\mathcal{C}^0)$, and $(\delta\mathcal{H}^{\vec{k}}, \delta\mathcal{C}^{\vec{k}})$ are, respectively, a pair of gauge-invariant variables (the so-called Dirac observables) and two pairs each made of a constraint and a conjugate variable that serve as a gauge-fixing function. The gauge-invariant variables are the only dynamical part of cosmological perturbations but have no unambiguous physical meaning. In order to reconstruct the physical spacetime we need to introduce a coordinate system by assuming gauge-fixing conditions $\delta\mathcal{C}^0=0$ and $\delta\mathcal{C}^{\vec{k}}=0$. The former fixes the foliation, whereas the latter introduces a threading in the perturbed spacetime. This is illustrated in Fig. \ref{coordinates}. 

In order to reconstruct the spacetime geometry, we invert the mapping \eqref{map} for $\delta\mathcal{H}^0=0$, $\delta\mathcal{H}^{\vec{k}}=0$, $\delta\mathcal{C}^0=0$, and $\delta\mathcal{C}^{\vec{k}}=0$. As a result, we obtain the ADM and the dust perturbation variables as functions of the gauge-invariant variables $\Psi$ and $\Pi_{\Psi}$,
$$\delta q_{ab}=\delta q_{ab}(\Psi, \Pi_{\Psi}),~~\delta \pi^{ab}=\delta \pi^{ab}(\Psi, \Pi_{\Psi}).$$ (For more details, see \cite{Boldrin_2022}.) Furthermore, the lapse and shift perturbations, $\delta N_{0}$ and $\delta N^{\vec{k}}$, are reconstructed from the stability condition for the gauge-fixing functions,
\begin{align}
\delta\dot{\mathcal{C}}^0=0,~~~\delta\dot{\mathcal{C}}^{\vec{k}}=0.
\end{align}

Making use of the gauge-fixing conditions $\delta\mathcal{C}^0=0$ and $\delta\mathcal{C}^{\vec{k}}=0$ and the linear constraints we reduce the Hamiltonian \eqref{hamADM} by reducing the number of dynamical variables to two. Then these two variables are transformed into a canonical pair $(\Psi, \Pi_{\Psi})$, where $\Psi$ is the Bardeen potential, a convenient and popular gauge-invariant perturbation variable. The reduced Hamiltonian that generates the dynamics of the Bardeen potential and its conjugate momentum is found to read
\begin{align}\label{hamred}
H_{phys}=\frac{N p^2}{192 a^3k^2}\Pi_{\Psi}^2,
\end{align}
where fixing $N=a$ yields the conformal time Hamiltonian. One may verify that this Hamiltonian yields the well-known dynamics of the Bardeen potential in the matter-dominated era, $\Psi_{,\eta\eta}+\frac{6}{\eta}\Psi_{,\eta}=0$.

\subsection{Gauge transformations}
It is obvious that the canonical transformation \eqref{map} is not unique. In particular, depending on the problem on hand, different sets of gauge-fixing functions $\delta\mathcal{C}^0$ and $\delta\mathcal{C}^{\vec{k}}$ may be most useful.  It can be shown that up to weakly vanishing terms any two sets of gauge-fixing functions differ by a combination of gauge-invariant variables; thus, in our case,
\[\delta \tilde{C}^{0}\approx\delta C^0 +\alpha^0 \Pi_{\Psi}+\beta^{0} \Psi,~~\delta \tilde{C}^{k}\approx\delta C^k +\alpha^k\Pi_{\Psi}+\beta^{k} \Psi,\]
where $(\alpha^0, \beta^{0}, \alpha^k, \beta^{k})$ are arbitrary background functions. The choice of the origin $(\delta C^0,\delta C^k)$ is also arbitrary. The above formula produces the complete space of gauge-fixing functions \cite{Boldrin_2022}. 

We emphasize that the lapse and shift perturbations are implied by the gauge-fixing conditions. Following the theory developed in \cite{Boldrin_2022}, the lapse and shift perturbations are changed upon the above gauge transformation by an amount that depends solely on the  coefficients $\alpha$'s and $\beta$'s, and the reduced Hamiltonian \eqref{hamred} but not the full Hamiltonian \eqref{hamADM}.

\subsection{Matching hypersurface}

Our goal is to sew at a fixed constant-time hypersurface a suitable spatial average of an inhomogeneous universe with a homogeneous Bianchi IX model. The geometry of the latter is given in a fixed reference system: (i) the flow of the fluid is orthogonal to the three-surfaces and (ii) it drags irrotationally a spatial basis from one three-surface to the next one. To avoid any abrupt change of the velocity of the fluid at the matching hypersurface, we fix such a gauge condition for the perturbed FLRW, in which the fluid's flow is orthogonal. Thus, we impose on cosmological perturbations
\begin{align}\label{gc1}
    \delta\phi=0,
\end{align} 
which foliates the spacetime with suitable (i.e., ``fluid-orthogonal") three-surfaces. Note that this foliation is mathematically globally valid though physically limited to the regime of the applicability of perturbation theory. This is our first gauge-fixing condition. 

We shall complete the gauge-fixing procedure by setting a coordinate system on each of the above three-surfaces. In what follows we assume that at the perturbation level our coordinates are free of singularities such as shell crossings that could, in principle, spoil our results. In the nonlinear regime, for describing finite domains of the universe we employ the spatially homogeneous metric of the BIX model in which shell crossings are absent.

\subsection{Basis in the matching hypersurface}\label{basis}

So far we have discussed the foliation of an inhomogeneous spacetime, which is compatible with the natural foliation of the Bianchi IX model. This corresponds to the first gauge-fixing condition $\delta\mathcal{C}^0=0$. We have not yet specified the spatial basis in which the three-metric and the three-momentum components should be fitted on the matching hypersurface.

The spatial geometry of the Bianchi IX model is expressed in nonholonomic basis made of 1-forms $\omega^a$ (and the dual vector $e_a$) such that
\begin{align}
\ud \omega^a=\frac{n}{2}\epsilon^a_{~bc}\omega^b\wedge\omega^c,
\end{align}
where $\epsilon^a_{~bc}$ is the total antisymmetric tensor with $\epsilon^1_{~23}=1$. Such 1-forms obviously exist independent of the metric and could be similarly constructed on the matching hypersurface of the perturbed flat Friedmann universe. Making use of these 1-forms for the perturbed universe is necessary since the best-fit three-metric and three-momentum must be diagonal in this particular basis under the condition that the averaged region is positively curved. The 1-forms {\it a priori} can be defined only up to global $SO(2)$ transformations. However, as we shall see below, once the perturbations are specified, the 1-forms in our construction become fully specified too. 

Let us consider the Bianchi IX 1-forms on the matching hypersurface. We introduce the transformation matrix $S_{ai}(y)$ such that
\begin{align}
\omega^a=S_{~i}^a(y)\ud y^i,
\end{align}
where $y^i$ for $i=1,2,3$ form a coordinates chart on the Bianchi IX hypersurface $\mathbb{S}^3$ in the neighborhood of the pole $(0,0,0)$ and confined to the region with $\zeta^2:=\frak{n}^{-2}-\sum_i(y^i)^2>0$. We set
\begin{widetext}
\begin{align}\begin{split}
\omega^1&=2\frak{n}\left[-\frac{(y^1)^2+\zeta^2}{\zeta}~\ud y^1-\left(y^3+\frac{y^2y^1}{\zeta}\right)\ud y^2+\left(y^2-\frac{y^3y^1}{\zeta}\right)\ud y^3\right],\\
\omega^2&=2\frak{n}\left[\left(y^3-\frac{y^2y^1}{\zeta}\right)\ud y^1-\frac{(y^2)^2+\zeta^2}{\zeta}~\ud y^2-\left(y^1+\frac{y^2y^3}{\zeta}\right)\ud y^3\right],\\
\omega^3&=2\frak{n}\left[-\left(y^2+\frac{y^3y^1}{\zeta}\right)\ud y^1+\left(y^1-\frac{y^2y^3}{\zeta}\right)\ud y^2-\frac{(y^3)^2+\zeta^2}{\zeta}~\ud y^3\right].
\end{split}
\end{align}
\end{widetext}
We shall now use these Bianchi IX 1-forms induced on the matching hypersurface to restate the perturbed three-metric and three-momentum in the new basis, namely,
\begin{align}\label{transf}\begin{split}
 \tilde{q}_{ab}(x,y)&=S_{~a}^{-1i}(y)S_{~b}^{-1j}(y) q_{ij}(x),\\
\tilde{\pi}^{ab}(x,y)&=S_{~i}^a(y)S_{~j}^b(y)\pi^{ij}(x).\end{split}
\end{align}
Next, we identify the Bianchi IX coordinate system $\{ y^i\}$ with the spatial coordinate system of the FRW model $\{x^i\}$ as they both are defined on the same hypersurface. The matching is geometrically consistent only if the Bianchi IX spatial curvature and the averaged spatial curvature of a given domain $\mathcal{D}$ within the perturbed Friedmann universe are equal; that is,
\begin{align}\label{matchR}
\frac{3\frak{n}^2_{\mathcal{D}}}{2e^{2\beta_0}}=\langle\delta R\rangle_\mathcal{D}, 
\end{align}
where $\langle\dots\rangle_\mathcal{D}$ denotes an average over $\mathcal{D}$, defined in Eq. \eqref{ave}. The curvature parameter $\frak{n}^2_{\mathcal{D}}$ is specific to the particular domain as emphasized by the notation. Note that $\mathcal{V}_{\mathcal{D}}=\frac{16\pi^2}{\frak{n}^3_{\mathcal{D}}}$ corresponds to the entire volume of the Bianchi IX model, and not to the volume of the matched domain $\mathcal{D}$, which must be much smaller. Upon these identifications, the formula \eqref{transf} becomes
\begin{align}\begin{split}
 \tilde{q}_{ab}(x)&=S_{~a}^{-1i}(x)S_{~b}^{-1j}(x) q_{ij}(x),\\
\tilde{\pi}^{ab}(x)&=S_{~i}^a(x)S_{~j}^b(x)\pi^{ij}(x),\end{split}
\end{align}
where the comoving coordinates $\{x^i\}$ are set in the usual way as the distances in megaparsecs comoving. 

\begin{figure}
    \centering
    \includegraphics[width=0.33\textwidth]{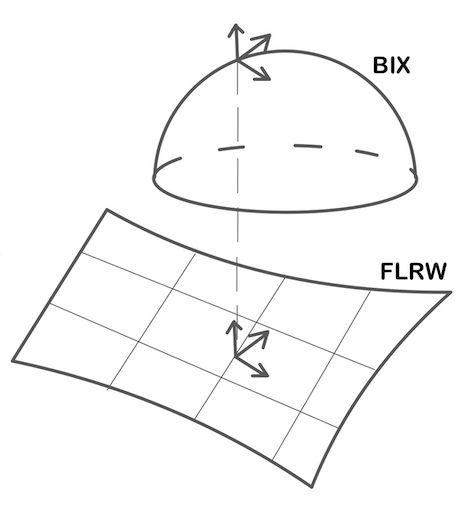}
    \caption{The three Bianchi IX vector fields are transported to a domain in the matching hypersurface where they are expressed in the Cartesian coordinates.}\label{mapp}
\end{figure}

Given that $\frak{n}_{\mathcal{D}}^{-2}-\sum_i(y^i)^2>0$, and hence $\frak{n}_{\mathcal{D}}^{-2}-\sum_i(x^i)^2>0$, we expand the forms $\omega^{a}$ and their duals $e_a$ in the coordinates $\{x^i\}$ around the pole $(0,0,0)$,
\begin{align}\label{invap}\begin{split}
\omega^a&\simeq- 2\ud x^a+2\epsilon^a_{~bc}(x^b\frak{n}_{\mathcal{D}})\ud x^c+(\vec{x}\frak{n}_{\mathcal{D}})^2\ud x^a\\
&-2(x^a\frak{n}_{\mathcal{D}})(x^b\frak{n}_{\mathcal{D}})\delta_{bc}\ud x^c,\\
e_a&\simeq- \frac12\partial_{x^a}-\frac12\epsilon^c_{~ba}(x^b\frak{n}_{\mathcal{D}})\partial_{x^c}+\frac14(\vec{x}\frak{n}_{\mathcal{D}})^2\partial_{x^a},
\end{split}
\end{align}
omitting all higher-order terms in $x\frak{n}_{\mathcal{D}}$. The latter is justified as the typical amplitude of curvature perturbations on the comoving hypersurfaces $\frac{4a^2}{k_{\mathcal{D}}^2}\langle\delta R\rangle_\mathcal{D}\approx\sqrt{\mathcal{P}_{\mathcal{R}}(k_{\mathcal{D}})}\approx\sqrt{2.1\cdot 10^{-9}}$ implies that the comoving size of the domain $k^{-1}_{\mathcal{D}}\approx \sqrt{\frac{\sqrt{\mathcal{P}_{\mathcal{R}}(k_{\mathcal{D}})}}{24}}\frak{n}^{-1}_{\mathcal{D}}$ is much smaller than the respective radius of the three-sphere of the Bianchi IX model. Thus, the invariant 1-forms and vectors \eqref{invap} are valid approximations on these domains. Figure \ref{mapp} depicts our construction of the BIX triad in the perturbed FLRW universe.

Expressing the three-geometry of the perturbed Friedmann universe in the new basis \eqref{invap}, we find (note that three-momentum is a densitized tensor) at zeroth order,
\begin{align}\label{0q}\begin{split}
\tilde{q}_{ab}=\frac14 a^2\delta_{ab},~~\tilde{\pi}^{ab}=\frac16 p \delta^{ab},
\end{split}
\end{align}
and at first order,
\begin{align}\label{tiltedgeo}\begin{split}
\delta \tilde{q}_{ab}(x)&\simeq \frac14\delta q_{ab}(x)-\frac{\frak{n}_{\mathcal{D}}^2}{4}a^2 x^ax^b,\\
\delta\tilde{\pi}^{ab}(x)&\simeq \frac12~ \delta\pi^{ab}(x)+\frac{\frak{n}_{\mathcal{D}}^2}{6}\left(x^ax^b+\frac{\vec{x}^2}{2}\right)p,\end{split}
\end{align}
where we omitted higher-order terms in $x^b\frak{n}_{\mathcal{D}}$. We note that the nondiagonal terms $\propto x^ax^b$ in the three-metric $\delta \tilde{q}_{ab}(x)$ cannot be removed by any rotation of the nonholonomic basis. However, these specific terms always vanish upon integration because of the reflective symmetry of the domain. We conclude that the three-metric tensor in the nonholonomic basis $\delta \tilde{q}_{ab}(x)$ becomes diagonal after averaging only if the coordinate metric $\delta q_{ab}(x)$ is diagonal, that is,
\begin{align}\label{gc2}
    \delta q_2=0.
\end{align}
This is our second and final gauge-fixing condition. A discussion on the relation between the two gauge-fixing conditions and the coordinate system at the matching surface and beyond is found in Appendix \ref{comment}.

\subsection{Bardeen potential on the matching hypersurface}

The matching to the homogeneous model is assumed during matter-dominated era in the absence of sound waves that could spoil the homogeneous approximation. During this era, the Bardeen potential rapidly becomes constant with $\Psi\gg\eta\acute{\Psi}$ at any scale. The three-metric and three-momentum are some functions of the Bardeen potential, which are specified upon choosing the gauge-fixing conditions \eqref{gc1} and \eqref{gc2}. We find 
\begin{align}\label{matmatch}\begin{split}
 \delta q_1&=-\frac{10}{3}a^2\Psi,~~\delta q_2=0,\\
 \delta\pi_1&=\left[-\frac53+8(k/p)^2\right]p\Psi,\\
  \delta\pi_2&=-\frac{8}{3}(k/p)^2p\Psi,\end{split}
\end{align}
where we assumed $\Psi\gg\eta\acute{\Psi}$. Note that for subhorizon modes, the factor $(k/p)^2\propto (k/\mathcal{H})^2$ enhances the three-momentum perturbation. We conclude that the three-metric and three-momentum perturbations in position representation and in the Bianchi IX-invariant basis read
 \begin{align}\label{hyperdata}
 \begin{split}
 \delta \tilde{q}_{ab}&=-\frac{5}{6}a^2\Psi~\delta_{ab}-\frac{\frak{n}_{\mathcal{D}}^2}{4}a^2~x^ax^b,\\
   \delta\tilde{\pi}^{ab}& =-\frac16\left[\frac53 p\Psi+\frac{12}{p} \Psi_{,cc}\right]~\delta^{ab}+2p^{-1}\Psi_{,ab}\\
   &+\frac{\frak{n}_{\mathcal{D}}^2}{6}\left(x^ax^b+\frac{\vec{x}^2}{2}\right)p
 \end{split}
 \end{align}
We note that on the matching hypersurface the shear is generically nonzero and given by $2p^{-1}\Psi_{,ab}$.

\section{\label{sec:average}Fitting procedure}

\subsection{Averaging}
For any perturbation variable $\delta{X}(x,\eta)$, we introduce its average,
\begin{align}\label{ave}
\langle\delta{X}\rangle_{\mathcal{D}}=\delta{X}_{\mathcal{D}}=\int_{\Sigma} \delta{X}(x,\eta)W_{\mathcal{D}}(x)\sqrt{-q}~\ud^3x,
\end{align}
where $W_{\mathcal{D}}(x)$ is a normalized window function that to a quantity $\delta{X}(x,\eta)$ associates its spatial average $\delta{X}_{\mathcal{D}}(\eta)$ over a domain $\mathcal{D}$ characterized by a length $L=2\pi k_{\mathcal{D}}^{-1}$ and $\sqrt{q}$ is the spatial metric density. Analogously, the average of a derivative of $\delta{X}(x,\eta)$ reads
\begin{align}
\langle\partial_a\delta{X}\rangle_{\mathcal{D}}=\delta{X}_{,a{\mathcal{D}}}=\int_{\mathcal{D}} \delta{X}(x,\eta)_{,a}W_{\mathcal{D}}(x) \sqrt{q}~\ud^3x,
\end{align}
with the averages of all the higher derivatives similarly defined.

We apply this averaging procedure to \eqref{hyperdata} and find
\begin{align}\label{match1}
 \begin{split}
 \delta \tilde{q}_{ab{\mathcal{D}}}&=\left[-\frac{5}{6}a^2\Psi_{\mathcal{D}}-\frac{\frak{n}_{\mathcal{D}}^2L^2}{20}a^2\right]\delta_{ab},\\
   \delta\tilde{\pi}^{ab}_{\mathcal{D}}&= -\frac16\left[\frac53 p\Psi_{\mathcal{D}}+\frac{12}{p} \Psi_{,cc{\mathcal{D}}}-\frac{\frak{n}_{\mathcal{D}}^2L^2}{2}p\right]\delta^{ab}\\
   &+2p^{-1}\Psi_{,ab{\mathcal{D}}},
 \end{split}
 \end{align}
where the components have been averaged over a ball of radius $L$ centered at $(0,0,0)$.

The averaged components of the three-metric and the three-momentum must be now simultaneously diagonalized in order to be matched to the diagonal Bianchi IX metric. The three-metric is already diagonal by virtue of the second gauge-fixing condition $\delta q_2=0$. The three-momentum commutes with the three-metric, as the latter is proportional to the identity matrix, and thus they can be simultaneously diagonalized. Let us assume a rotation matrix of the form $R\simeq\mathbf{1}+\delta R$ with $\delta R$ being a small perturbation, which diagonalizes the averaged three-momentum $\frac43p\delta^{ab}+\delta\tilde{\pi}^{ab}_{\mathcal{D}}$,
\begin{align}
(\mathbf{1}+\delta R)(\frac43 p\delta^{ab}+\delta\tilde{\pi}^{ab}_{\mathcal{D}})(\mathbf{1}-\delta R)\simeq \frac43 p\delta^{ab}+\delta\tilde{\pi}^{ab}_{\mathcal{D}},
\end{align}
since the first-order term $[\delta R,\frac43 p\delta^{ab}]$ vanishes identically. Hence, the first-order eigenvalues of the three-momentum are $\delta\tilde{\pi}^{11}_{\mathcal{D}}$, $\delta\tilde{\pi}^{22}_{\mathcal{D}}$, and $\delta\tilde{\pi}^{33}_{\mathcal{D}}$, where
\begin{align*}\begin{split}\delta\tilde{\pi}^{(aa)}_{\mathcal{D}} &= -\frac{5}{18} p\Psi_{\mathcal{D}}-2p^{-1} \Psi_{,cc{\mathcal{D}}}+2p^{-1}\Psi_{,(aa){\mathcal{D}}}\\
   &+\frac{\frak{n}_{\mathcal{D}}^2L^2}{12}p,\end{split}\end{align*}
where the indices in the parentheses are not summed over. In general, the matrix $\delta R$ is not small and we need to determine the eigenvalues of the full averaged three-momentum.

\subsection{Matching}
Upon combining Eqs. \eqref{pertb9}, \eqref{0q}, and \eqref{match1}, we arrive at the following relations at the matching hypersurface:
\begin{align}\begin{split}\label{matchfinal}
&\delta\beta_0=-\frac{5}{3} \Psi_{\mathcal{D}}-\frac{L^2}{10}\frak{n}_{\mathcal{D}}^2,~~\delta\beta_+=0,~~\delta\beta_-=0,\\
&\delta p_0=\frac{a^2(p^2(3\frak{n}_{\mathcal{D}}^2L^2-50\Psi_{\mathcal{D}}) -80\Psi_{,cc{\mathcal{D}}})}{40 p},\\
&\delta p_+=a^2p^{-1}\left(\Psi_{,11{\mathcal{D}}}+\Psi_{,22{\mathcal{D}}}-2\Psi_{,33{\mathcal{D}}}\right),\\
&\delta p_-=\sqrt{3}a^2p^{-1}\left(\Psi_{,11{\mathcal{D}}}-\Psi_{,22{\mathcal{D}}}\right),
\end{split}\end{align}
while the background quantities are related as follows:
\begin{align}\begin{split}
e^{\beta_0}=\frac{1}{2}a, ~~p_0=\frac14a^2p,~~\frak{n}^2_{\mathcal{D}}=\frac{10}{9}\Psi_{,cc{\mathcal{D}}},
\end{split}\end{align}
where we used the results of Sec. \ref{basis} and the fact that $\delta R=\frac{20}{3a^2}\Psi_{,cc}$ in the gauge \eqref{arbitrarygauge} (note that $p=-6\dot{a}$). Using Eqs \eqref{matchfinal}, we have checked that the averages of the scalar quantities $\delta R$, $\delta V$, $\delta\rho$, and $\delta\theta$ are identical to the corresponding Bianchi IX quantities at the matching surface.

\begin{figure}
    \centering
    \includegraphics[width=0.5\textwidth]{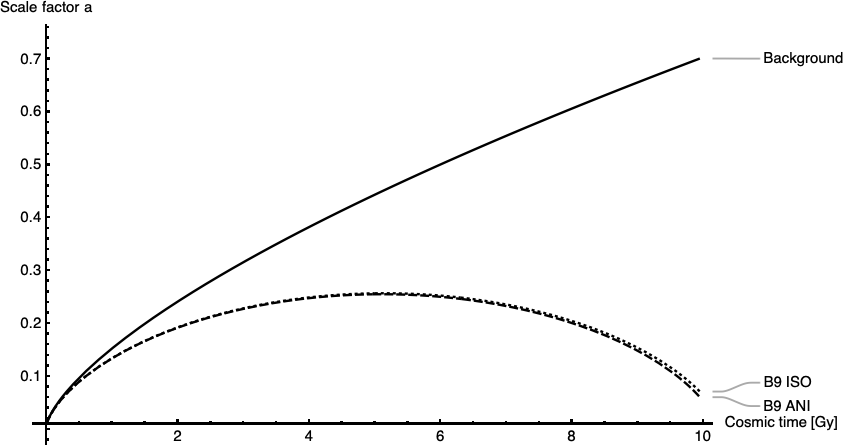}
    \vspace{0.2cm}

    \includegraphics[width=0.48\textwidth]{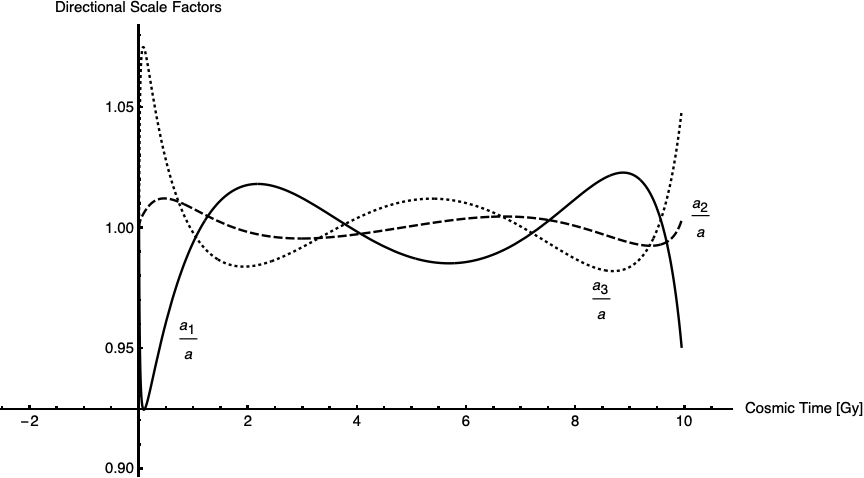}
     \vspace{0.2cm}
    
    \includegraphics[width=0.48\textwidth]{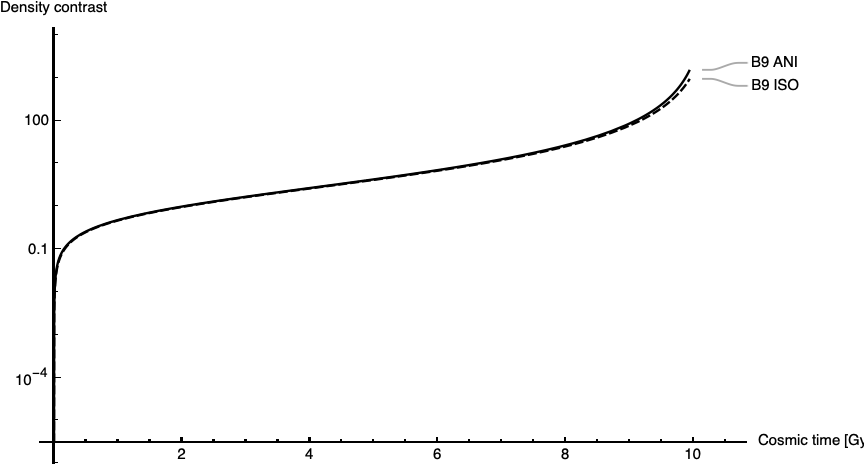}
    \caption{The behavior of the anisotropic and isotropic Bianchi IX models in comparison to the background evolution for the first initial dataset.}\label{typical}
\end{figure}

\subsection{\label{sec:examples}Numerical examples}

Let us illustrate our method with examples of dynamics of two specific domains: one with $L=10$Mpc and the other with $L=50$Mpc (comoving). We choose the Einstein--de Sitter background for convenience. We set the initial conditions for the domains at a redshift $z=100$. For the first domain, we set the initial density contrast $\delta=0.023642$, corresponding to the variance of the probability distribution for $\delta$ at $z=100$ and at a scale of $10$Mpc. The domain-averaged values for the Bardeen potential and its derivative are as follows: 
$$\Psi_{\mathcal{D}}=1.85615\cdot 10^{-7},~\Psi_{,11\mathcal{D}}=4.11261\cdot 10^{-7},$$
$$\Psi_{,22\mathcal{D}}=0,~ \Psi_{,33\mathcal{D}}=-3.5251\cdot 10^{-7}.$$ For the other domain, we set the initial density contrast $\delta=0.011073$, corresponding to the variance of the probability distribution for $\delta$ at $z=100$ and at a scale of $50$Mpc. The domain-averaged values for the Bardeen potential and its derivative are as follows: 
$$\Psi_{\mathcal{D}}=1.88899\cdot 10^{-6},~\Psi_{,11\mathcal{D}}=1.92619\cdot 10^{-7},$$
$$\Psi_{,22\mathcal{D}}=0,~\Psi_{,33\mathcal{D}}=-1.65102\cdot 10^{-7}.$$

In Fig. \ref{typical} (upper panel) the evolution of the scale factor of the 10 Mpc domain is plotted. The domain undergoes a turnaround at redshift $z=1.224$ and then collapses (i.e., reaches the critical density contrast of $177$) at redshift $z=0.485$. For comparison, in the same figure, the evolution of this domain, starting from the same initial condition except for the fact that the initial shear is neglected is plotted. There is a slight but visible difference in the evolution between the spherical and the nonspherical model, in particular, toward the collapse. The nonspherical model collapses first due to the nonvanishing shear that slows down expansion and fuels contraction. The turnaround and the collapse redshifts in the spherical model are, respectively, $z=1.213$ and $z=0.477$. In Figure \ref{typical} (middle panel), we can see the evolution of the directional scale factors that exhibit changes in the shape of the domain at the level of a few percents. Fig. \ref{typical} (lower panel) presents the evolution of the density contrast in cosmic time, where the differences between the two models are again small but noticeable. These results do not exhibit a significant departure from the spherical collapse model, however, this is due to the specific initial conditions chosen. The broad-based influence of the shear on the abundance of the collapsed objects would be revealed through the statistical examination i.e. by the mass function of the galaxy clusters.

\begin{figure}
    \centering
     \includegraphics[width=0.5\textwidth]{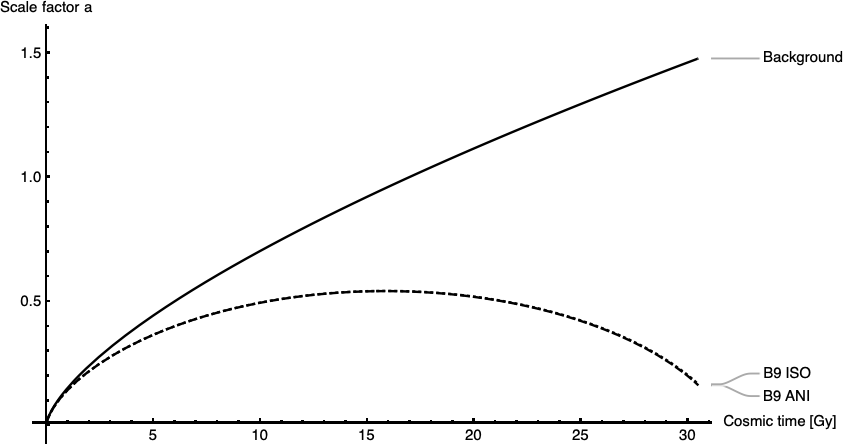}
    \vspace{0.2cm}
    
    \includegraphics[width=0.48\textwidth]{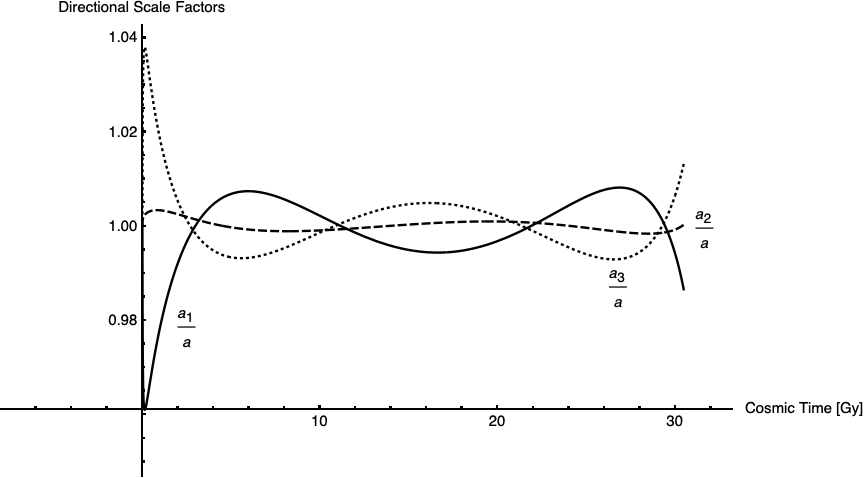}
     \vspace{0.2cm}
    
    \includegraphics[width=0.48\textwidth]{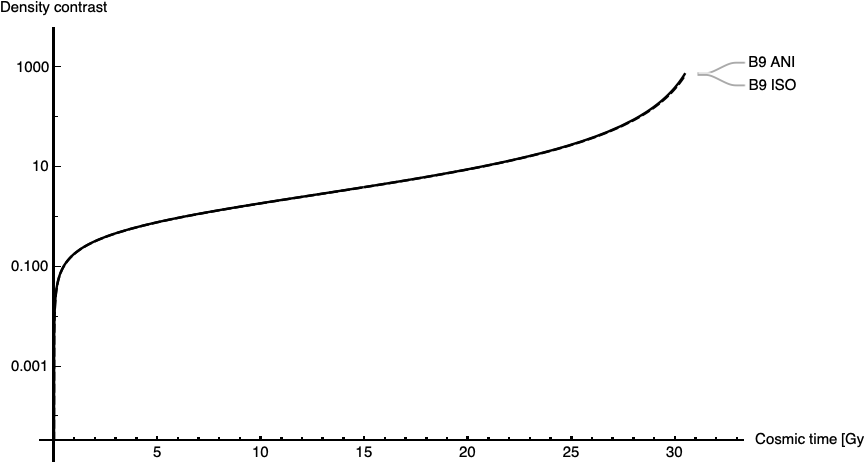}
    \caption{The behavior of the anisotropic and isotropic Bianchi IX models in comparison to the background evolution the second initial dataset.}\label{typical2}
\end{figure}

In Fig. \ref{typical2} (upper panel) the evolution of the scale factor of the 50 Mpc domain is plotted. The domain undergoes a turnaround at redshift $z=0.046$ and then collapses at redshift $z=-0.301$. There is no visible difference in the evolution between the spherical and the nonspherical model at this scale. The turnaround and the collapse redshifts in the spherical model are, respectively, $z=0.045$ and $z=-0.302$, so the difference with the nonspherical case occurs in the fourth significant figure. In Fig. \ref{typical2} (middle panel) we can see the behavior of the directional scale factors that exhibit changes in the shape of the domain at the level of 1\% for most of the evolution. Figure \ref{typical2} (lower panel) presents the evolution of the density contrast in cosmic time, where no difference between the two models is noticeable.

\section{\label{sec:level1}Summary}

In this work, we have formulated an anisotropic collapse model applicable to generic, positively curved domains of perturbed FLRW. This model incorporates the effects of anisotropy and curvature on the evolution of the collapsing domain in a relativistic manner. The Bianchi IX spacetime is the only anisotropic and positively curved homogeneous model that undergoes recollapse. 

Our key contribution lies in identifying a geometrically motivated method for defining a matching hypersurface (and a reference system on it) for the best fit of the Bianchi IX model.  Making use of the BIX symmetries as a guiding tool, we were able to sew a generic domain to a hypersurface-orthogonal Bianchi IX metric. This novel approach allows us to approximate a wide range of realistic cosmological scenarios with a spatially homogeneous, exact solution, requiring only the knowledge of initial power spectrum. The resulting spacetime model can approximately track evolution of a generic domain to an arbitrarily nonlinear regime.

Given the richness of the BIX-type models, our methodology has the potential for extension to account for other physical phenomena such as rotation, pressure, or a tilt in the motion of the fluid. 

As an immediate application, we have investigated the anisotropic dynamics of domains with density perturbations of a typical size. The shear present in this model decelerates the initial expansion, leading to shorter turnaround and virialization times. Consequently, the structure formation process appears to occur more rapidly — a well-known observation consistent with other independent approaches (see, for example, \cite{ShethTormen}).

It is important to emphasise that the idea of fitting the Bianchi models to inhomogeneous spacetimes in the context of averaging is not new. Notably, in the work by Spero and Baierlein \cite{1977Spero}, a method for determining the best-fit Bianchi metric is introduced. The authors employ a triad on an arbitrary spatial domain of the perturbed FLRW, respecting the domain's geometry. The structure coefficients of the triad are then employed to determine the structure constants of the best-fit Bianchi model. Using a variational approach, the authors reduce the problem of finding the most suitable structure constants to a set of differential equations with a boundary condition. Once solved, the best-fit Bianchi type is established. To identify the best-fit metric, the authors search within the family of all triads with the obtained structure constants to find the one that best reproduces the properties of the real metric.


While promising, the method introduced in \cite{1977Spero} has several problems that are circumvented by our approach. First, the obtained differential equations are difficult to solve even when the original metric is explicitly known, let alone a generic case. Second, the authors focus solely on fitting of a three-metric, leaving the issue of the second fundamental form unspecified. In contrast, our approach addresses these challenges more effectively. For the first issue, our reliance on the BIX as the only metric capable of describing the recollapse of the structure, thanks to positive curvature, simplifies the determination of best-fit structure constants. Moreover, since our procedure aims at generic initial conditions within positively curved domains, we adopt a more practical approach to specifying the explicit form of the best-fit BIX. As noted in \cite{1977Spero}, once the best-fit structure constants are determined, only space-independent rotations of the basis can be performed in search for the best-fit metric. As mentioned in the previous section, no rotations of nonholonomic basis that would keep the metric diagonal exist. However, we can utilize the fact that our procedure relies on fitting of BIX on a fixed hypersurface, i.e., our integration and transformation matrices are time independent. 

Among potential applications of our procedure, which we leave for future investigations, we include the formation of primordial black holes, the formation of virialized structures such as galaxies and galaxy clusters, and the examination of light propagation in the context of strong and weak lensing, as well as cosmological tests (see, for instance, \cite{JOIDG} for a cosmological test involving averaging techniques). In particular, the explicit scale-dependent metric provided by our method allows for a detailed examination of the lightlike structure of collapsing domains. This feature proves to be an indispensable tool in lensing analysis and apparent horizon estimations. It represents a substantial advantage compared to purely scalar-based approximation schemes that are unable to reconstruct the geometrical details of a given collapsing object. 

\begin{acknowledgments}
P. M. acknowledges the support of the National Science Centre
(NCN, Poland) under the research Grant No. 2018/30/E/ST2/00370. J. J. O.
and I. D. G. acknowledge the support of the National Science Centre (NCN, Poland) under the Sonata-15 research
Grant No. 2019/35/D/ST9/00342.
\end{acknowledgments}

\appendix

\section{Hamiltonian formalism for cosmological perturbation theory}\label{adm}
The second-order scalar constraint reads as follows:
\begin{align}\begin{split}
\mathcal{H}^{(2)}&=a\delta\pi_{ab}\delta\pi^{ab}-\frac{1}{2}a(\delta\pi)^2+\frac{a^{-1}p}{3}\delta\pi^{ab}\delta q_{ab}\\
&-\frac{a^{-1}p}{6}\delta\pi\delta q+\frac{5a^{-3}p^2}{72}\delta q_{ab}\delta q^{ab}+\frac{7a^{-3}p^2}{48}(\delta q)^2\\
&+\frac{a^{-3}}{2}\left(\delta q_{ab,ab}\delta q+\frac{3}{2}\delta q_{aa,c}\delta q_{bb,c}-\delta q_{ab,b}\delta q_{ac,c}\right.\\
&\left.+\frac{1}{2}\delta q_{ab,c}\delta q_{ab,c}\right)+\frac{a^{-2}\pi_{\phi}}{2}\delta^{ab}\delta\phi_{,a}\delta\phi_{,b}~,
\end{split}
\end{align}
while the linearized scalar and vector constraints read
\begin{align}\label{consA}\begin{split}
\delta\mathcal{H}^0&=-\frac{ap}{3}\delta\pi-\frac{a^{-1}p^2}{36}\delta q-a^{-1}(\delta q_{ab,ab}\\
&-\delta q_{aa,bb})+{\delta \pi_{\phi}}~,\\
\delta\mathcal{H}^i&=-2(a^2\delta\pi^{ij}_{,j}+\frac{1}{3}p\delta q_{ij,j})+\pi_{\phi}\delta\phi_{,i}~.\end{split}
\end{align}

\section{Perturbation variables}\label{kuch}
The Bardeen potential $\Psi$ and its conjugate momentum $\Pi_{\Psi}$ can be expressed in terms of the ADM variables as follows (in the momentum space):
\begin{align*}
    \Psi&=\frac{-4 a^2 p\delta\pi_1 + 2 p^2\delta q_1-8k^2(3 \delta q_1-\delta q_2)+a p^3 \delta{\phi}}{48 a^2 k^2},\\
    \Pi_{\Psi}&= -\frac{10 a^2}{3} \delta\pi_1+\left(4 k^2 +\frac{5p^2}{6}\right)a~ \delta \phi -\frac{8k^2}{3p} (3 \delta q_1\\
    &-\delta q_2)+\frac{5p}{3} \delta q_1.
\end{align*}
The canonical pair $(\Psi,\Pi_{\Psi})$, combined with the canonical pairs $(\delta\mathcal{H}^0,\delta{C}^{0})$,  $(\delta\mathcal{H}^{k},\delta {C}^{k})$ defined by Eqs \eqref{consA} and \eqref{arbitrarygauge}, form a complete canonical parametrization of the ADM perturbation phase space.

\section{Discussion on gauge fixing \label{comment}}

The assumed gauge-fixing conditions \eqref{gc1} and \eqref{gc2} imply the following canonical gauge-fixing functions:
\begin{align}\label{arbitrarygauge}\begin{split}
\delta{C}^{0}&= - \delta \phi,\\
\delta {C}^{k}&= -\frac{i \delta q_2}{2a^2k}.\end{split}\end{align}
It is important to note that this choice of gauge implies $\delta N=0$ and $\delta N^k=\frac{i4k}{ap}\Psi$. Thus, the flow of time (contrary to the flow of dust) is not orthogonal to the matching surface. This means that the dust mass is not conserved in this coordinate system. However, it is irrelevant for the outcome of the matching procedure as the latter is performed on a fixed constant-time slice within a given domain. Once the mass crosses the matching surface, its subsequent evolution is described by the Bianchi IX model in which the mass is conserved. 

Note that at the matching hypersurface one could modify the gauge-fixing function $\delta {C}^{k}$ as follows:
\begin{align}\begin{split}
\delta {C}^{k}= -\frac{i \delta q_2}{2a^2k} +\alpha^k\Pi_{\Psi}+\beta^{k} \Psi,\end{split}\end{align}
where $\alpha^k$ and $\beta^{k}$ are the previously discussed background functions that encode the freedom in choosing a coordinate systems on three-surfaces, including the matching surface. In \cite{Boldrin_2022}, a Hamiltonian theory of gauge-fixing and spacetime reconstruction for cosmological perturbations was developed. According to this theory, the spatial coordinates flow with the dust, i.e., the shift function vanishes, only when the gauge-fixing function $\delta {C}^{k}$ is modified with $\alpha^k$ and ${\beta}^k$ such that (we use conformal time)
\begin{align}
\acute{\alpha}^k=-\frac{p^2}{192 a^2k^2}\beta^k,~~\acute{\beta}^k=-\frac{i4k}{ap}.
\end{align}
The above coefficients come from the value of the shift function $\delta N^k=\frac{i4k}{ap}\Psi$ and the physical Hamiltonian \eqref{hamred}. The above equation implies that setting $\alpha^k=0$, $\beta^{k}=0$, $\acute{\alpha}^k=0$, and $\acute{\beta}^k=-\frac{i4k}{ap}$ ensures the vanishing of the shift function at the hypersurface while keeping the instantaneous gauge-fixing function $\delta {C}^{k}$ unaltered. Consequently, both the instantaneous diagonality of the metric and mass conservation are ensured. However, as time proceeds $\alpha^k$ and $\beta^{k}$ necessarily evolve and the domain becomes anisotropic with $\delta q_2\neq 0$. In other words, it is impossible to find a global coordinate system in which both $\delta q_2= 0$ and $\delta N^k=0$. Thus, for each hypersurface we need to impose the spatial coordinate system and its infinitesimal time development independent of the neighboring hypersurfaces. The described construction does not lead to global threads in the spacetime but this is not necessary; the only necessary property is that the matching hypersurfaces foliate globally the spacetime and the latter is satisfied thanks to the first gauge-fixing condition $\delta {C}^{0}=0$ that holds globally. This construction is merely an illustration that a continuous transfer of the spatial coordinates through a matching surface and aligned with the dust flow exists. However, this fact has no physical consequences, as neither the three-metric nor the three-momentum on a given hypersurface depends on how the spatial coordinate system develops beyond this hypersurface. 

Given a specified coordinate system $\{x^i\}$ on a well-defined matching surface $t=const.$, the obtained Bianchi IX triad $\{e_a\}$ and its dual $\{\omega^a\}$ become uniquely defined. Hence, the formula \eqref{tiltedgeo} provides the unambiguous three-metric and three-momentum components to be averaged. Our approach insists on the use of the Bianchi IX basis, which makes the averaging process tied to the assumed symmetry group. The outcome of the best-fit procedure may differ when applied to a different time slice. We consider this slice dependence to be natural, with the approximation tending to be more accurate when applied at later times.

\bibliography{references}

\providecommand{\noopsort}[1]{}\providecommand{\singleletter}[1]{#1}%
\begin{thebibliography}{17}%
\makeatletter
\providecommand \@ifxundefined [1]{%
 \@ifx{#1\undefined}
}%
\providecommand \@ifnum [1]{%
 \ifnum #1\expandafter \@firstoftwo
 \else \expandafter \@secondoftwo
 \fi
}%
\providecommand \@ifx [1]{%
 \ifx #1\expandafter \@firstoftwo
 \else \expandafter \@secondoftwo
 \fi
}%
\providecommand \natexlab [1]{#1}%
\providecommand \enquote  [1]{``#1''}%
\providecommand \bibnamefont  [1]{#1}%
\providecommand \bibfnamefont [1]{#1}%
\providecommand \citenamefont [1]{#1}%
\providecommand \href@noop [0]{\@secondoftwo}%
\providecommand \href [0]{\begingroup \@sanitize@url \@href}%
\providecommand \@href[1]{\@@startlink{#1}\@@href}%
\providecommand \@@href[1]{\endgroup#1\@@endlink}%
\providecommand \@sanitize@url [0]{\catcode `\\12\catcode `\$12\catcode
  `\&12\catcode `\#12\catcode `\^12\catcode `\_12\catcode `\%12\relax}%
\providecommand \@@startlink[1]{}%
\providecommand \@@endlink[0]{}%
\providecommand \url  [0]{\begingroup\@sanitize@url \@url }%
\providecommand \@url [1]{\endgroup\@href {#1}{\urlprefix }}%
\providecommand \urlprefix  [0]{URL }%
\providecommand \Eprint [0]{\href }%
\providecommand \doibase [0]{https://doi.org/}%
\providecommand \selectlanguage [0]{\@gobble}%
\providecommand \bibinfo  [0]{\@secondoftwo}%
\providecommand \bibfield  [0]{\@secondoftwo}%
\providecommand \translation [1]{[#1]}%
\providecommand \BibitemOpen [0]{}%
\providecommand \bibitemStop [0]{}%
\providecommand \bibitemNoStop [0]{.\EOS\space}%
\providecommand \EOS [0]{\spacefactor3000\relax}%
\providecommand \BibitemShut  [1]{\csname bibitem#1\endcsname}%
\let\auto@bib@innerbib\@empty
\bibitem [{\citenamefont {{Ellis}}\ and\ \citenamefont
  {{Stoeger}}(1987)}]{1987Ellis}%
  \BibitemOpen
  \bibfield  {author} {\bibinfo {author} {\bibfnamefont {G.~F.~R.}\
  \bibnamefont {{Ellis}}}\ and\ \bibinfo {author} {\bibfnamefont
  {W.}~\bibnamefont {{Stoeger}}},\ }\bibfield  {title} {\bibinfo {title} {{The
  'fitting problem' in cosmology}},\ }\href
  {https://doi.org/10.1088/0264-9381/4/6/025} {\bibfield  {journal} {\bibinfo
  {journal} {Classical and Quantum Gravity}\ }\textbf {\bibinfo {volume} {4}},\
  \bibinfo {pages} {1697} (\bibinfo {year} {1987})}\BibitemShut {NoStop}%
\bibitem [{\citenamefont {{Buchert}}(2000)}]{Buchert1}%
  \BibitemOpen
  \bibfield  {author} {\bibinfo {author} {\bibfnamefont {T.}~\bibnamefont
  {{Buchert}}},\ }\bibfield  {title} {\bibinfo {title} {{On Average Properties
  of Inhomogeneous Fluids in General Relativity: Dust Cosmologies}},\ }\href
  {https://doi.org/10.1023/A:1001800617177} {\bibfield  {journal} {\bibinfo
  {journal} {General Relativity and Gravitation}\ }\textbf {\bibinfo {volume}
  {32}},\ \bibinfo {pages} {105} (\bibinfo {year} {2000})},\ \Eprint
  {https://arxiv.org/abs/gr-qc/9906015} {arXiv:gr-qc/9906015 [gr-qc]}
  \BibitemShut {NoStop}%
\bibitem [{\citenamefont {{Buchert}}(2001)}]{Buchert2}%
  \BibitemOpen
  \bibfield  {author} {\bibinfo {author} {\bibfnamefont {T.}~\bibnamefont
  {{Buchert}}},\ }\bibfield  {title} {\bibinfo {title} {{On Average Properties
  of Inhomogeneous Fluids in General Relativity: Perfect Fluid Cosmologies}},\
  }\href {https://doi.org/10.1023/A:1012061725841} {\bibfield  {journal}
  {\bibinfo  {journal} {General Relativity and Gravitation}\ }\textbf {\bibinfo
  {volume} {33}},\ \bibinfo {pages} {1381} (\bibinfo {year} {2001})},\ \Eprint
  {https://arxiv.org/abs/gr-qc/0102049} {arXiv:gr-qc/0102049 [gr-qc]}
  \BibitemShut {NoStop}%
\bibitem [{\citenamefont {Jantzen}(2001)}]{Jantzen:2001me}%
  \BibitemOpen
  \bibfield  {author} {\bibinfo {author} {\bibfnamefont {R.~T.}\ \bibnamefont
  {Jantzen}},\ }\bibfield  {title} {\bibinfo {title} {{Spatially homogeneous
  dynamics: A Unified picture}},\ }\href@noop {} {\  (\bibinfo {year}
  {2001})},\ \Eprint {https://arxiv.org/abs/gr-qc/0102035}
  {arXiv:gr-qc/0102035} \BibitemShut {NoStop}%
\bibitem [{Note1()}]{Note1}%
  \BibitemOpen
  \bibinfo {note} {It is worth noting that the Bianchi models do not exhaust
  all the possible homogeneous cosmologies: the Kantowski-Sachs model is a
  homogeneous cosmology invariant under the group $\protect \mathbf {R}\times
  SO(3)$ that involves four independent parameters and does not posses any
  three-dimensional subgroup.}\BibitemShut {Stop}%
\bibitem [{\citenamefont {{Spero}}\ and\ \citenamefont
  {{Baierlein}}(1977)}]{1977Spero}%
  \BibitemOpen
  \bibfield  {author} {\bibinfo {author} {\bibfnamefont {A.}~\bibnamefont
  {{Spero}}}\ and\ \bibinfo {author} {\bibfnamefont {R.}~\bibnamefont
  {{Baierlein}}},\ }\bibfield  {title} {\bibinfo {title} {{Approximate symmetry
  groups of inhomogeneous metrics}},\ }\href {https://doi.org/10.1063/1.523425}
  {\bibfield  {journal} {\bibinfo  {journal} {Journal of Mathematical Physics}\
  }\textbf {\bibinfo {volume} {18}},\ \bibinfo {pages} {1330} (\bibinfo {year}
  {1977})}\BibitemShut {NoStop}%
\bibitem [{\citenamefont {{Spero}}\ and\ \citenamefont
  {{Baierlein}}(1978)}]{1978Spero}%
  \BibitemOpen
  \bibfield  {author} {\bibinfo {author} {\bibfnamefont {A.}~\bibnamefont
  {{Spero}}}\ and\ \bibinfo {author} {\bibfnamefont {R.}~\bibnamefont
  {{Baierlein}}},\ }\bibfield  {title} {\bibinfo {title} {{Approximate symmetry
  groups of inhomogeneous metrics: Examples}},\ }\href
  {https://doi.org/10.1063/1.523830} {\bibfield  {journal} {\bibinfo  {journal}
  {Journal of Mathematical Physics}\ }\textbf {\bibinfo {volume} {19}},\
  \bibinfo {pages} {1324} (\bibinfo {year} {1978})}\BibitemShut {NoStop}%
\bibitem [{\citenamefont {{Roukema}}\ and\ \citenamefont
  {{Ostrowski}}(2019)}]{2019Roukema}%
  \BibitemOpen
  \bibfield  {author} {\bibinfo {author} {\bibfnamefont {B.~F.}\ \bibnamefont
  {{Roukema}}}\ and\ \bibinfo {author} {\bibfnamefont {J.~J.}\ \bibnamefont
  {{Ostrowski}}},\ }\bibfield  {title} {\bibinfo {title} {{Does spatial
  flatness forbid the turnaround epoch of collapsing structures?}},\ }\href
  {https://doi.org/10.1088/1475-7516/2019/12/049} {\bibfield  {journal}
  {\bibinfo  {journal} {\jcap}\ }\textbf {\bibinfo {volume} {2019}},\ \bibinfo
  {eid} {049} (\bibinfo {year} {2019})},\ \Eprint
  {https://arxiv.org/abs/1902.09064} {arXiv:1902.09064 [astro-ph.CO]}
  \BibitemShut {NoStop}%
\bibitem [{\citenamefont {{Giani}}\ \emph {et~al.}(2022)\citenamefont
  {{Giani}}, \citenamefont {{Piattella}},\ and\ \citenamefont
  {{Kamenshchik}}}]{Giani}%
  \BibitemOpen
  \bibfield  {author} {\bibinfo {author} {\bibfnamefont {L.}~\bibnamefont
  {{Giani}}}, \bibinfo {author} {\bibfnamefont {O.~F.}\ \bibnamefont
  {{Piattella}}},\ and\ \bibinfo {author} {\bibfnamefont {A.~Y.}\ \bibnamefont
  {{Kamenshchik}}},\ }\bibfield  {title} {\bibinfo {title} {{Bianchi IX
  gravitational collapse of matter inhomogeneities}},\ }\href
  {https://doi.org/10.1088/1475-7516/2022/03/028} {\bibfield  {journal}
  {\bibinfo  {journal} {\jcap}\ }\textbf {\bibinfo {volume} {2022}},\ \bibinfo
  {eid} {028} (\bibinfo {year} {2022})},\ \Eprint
  {https://arxiv.org/abs/2112.01869} {arXiv:2112.01869 [gr-qc]} \BibitemShut
  {NoStop}%
\bibitem [{\citenamefont {Misner}(1969)}]{PhysRevLett.22.1071}%
  \BibitemOpen
  \bibfield  {author} {\bibinfo {author} {\bibfnamefont {C.~W.}\ \bibnamefont
  {Misner}},\ }\bibfield  {title} {\bibinfo {title} {Mixmaster universe},\
  }\href {https://doi.org/10.1103/PhysRevLett.22.1071} {\bibfield  {journal}
  {\bibinfo  {journal} {Phys. Rev. Lett.}\ }\textbf {\bibinfo {volume} {22}},\
  \bibinfo {pages} {1071} (\bibinfo {year} {1969})}\BibitemShut {NoStop}%
\bibitem [{\citenamefont {Ellis}\ and\ \citenamefont
  {MacCallum}(1969)}]{cmp/1103841345}%
  \BibitemOpen
  \bibfield  {author} {\bibinfo {author} {\bibfnamefont {G.~F.~R.}\
  \bibnamefont {Ellis}}\ and\ \bibinfo {author} {\bibfnamefont {M.~A.~H.}\
  \bibnamefont {MacCallum}},\ }\bibfield  {title} {\bibinfo {title} {{A class
  of homogeneous cosmological models}},\ }\href@noop {} {\bibfield  {journal}
  {\bibinfo  {journal} {Communications in Mathematical Physics}\ }\textbf
  {\bibinfo {volume} {12}},\ \bibinfo {pages} {108 } (\bibinfo {year}
  {1969})}\BibitemShut {NoStop}%
\bibitem [{\citenamefont {Uggla}(1997)}]{Uggla1997DynamicalSI}%
  \BibitemOpen
  \bibfield  {author} {\bibinfo {author} {\bibfnamefont {C.}~\bibnamefont
  {Uggla}},\ }\bibfield  {title} {\bibinfo {title} {Hamiltonian cosmology, in:
  Wainwright {J}., {E}llis {G}.{F}.{R}., eds. {D}ynamical {S}ystems in
  {C}osmology}\ }(\bibinfo  {publisher} {Cambridge University Press},\ \bibinfo
  {year} {1997})\BibitemShut {NoStop}%
\bibitem [{\citenamefont {Belinskii}\ \emph {et~al.}(1982)\citenamefont
  {Belinskii}, \citenamefont {Khalatnikov},\ and\ \citenamefont
  {Lifshitz}}]{doi:10.1080/00018738200101428}%
  \BibitemOpen
  \bibfield  {author} {\bibinfo {author} {\bibfnamefont {V.}~\bibnamefont
  {Belinskii}}, \bibinfo {author} {\bibfnamefont {I.}~\bibnamefont
  {Khalatnikov}},\ and\ \bibinfo {author} {\bibfnamefont {E.}~\bibnamefont
  {Lifshitz}},\ }\bibfield  {title} {\bibinfo {title} {A general solution of
  the einstein equations with a time singularity},\ }\href
  {https://doi.org/10.1080/00018738200101428} {\bibfield  {journal} {\bibinfo
  {journal} {Advances in Physics}\ }\textbf {\bibinfo {volume} {31}},\ \bibinfo
  {pages} {639} (\bibinfo {year} {1982})},\ \Eprint
  {https://arxiv.org/abs/https://doi.org/10.1080/00018738200101428}
  {https://doi.org/10.1080/00018738200101428} \BibitemShut {NoStop}%
\bibitem [{\citenamefont {Reiterer}\ and\ \citenamefont
  {Trubowitz}(2010)}]{reiterer2010bkl}%
  \BibitemOpen
  \bibfield  {author} {\bibinfo {author} {\bibfnamefont {M.}~\bibnamefont
  {Reiterer}}\ and\ \bibinfo {author} {\bibfnamefont {E.}~\bibnamefont
  {Trubowitz}},\ }\bibfield  {title} {\bibinfo {title} {The bkl conjectures for
  spatially homogeneous spacetimes},\ }\href@noop {} {\  (\bibinfo {year}
  {2010})},\ \Eprint {https://arxiv.org/abs/1005.4908} {arXiv:1005.4908
  [gr-qc]} \BibitemShut {NoStop}%
\bibitem [{\citenamefont {Boldrin}\ and\ \citenamefont
  {Ma{\l}kiewicz}(2022)}]{Boldrin_2022}%
  \BibitemOpen
  \bibfield  {author} {\bibinfo {author} {\bibfnamefont {A.}~\bibnamefont
  {Boldrin}}\ and\ \bibinfo {author} {\bibfnamefont {P.}~\bibnamefont
  {Ma{\l}kiewicz}},\ }\bibfield  {title} {\bibinfo {title} {Gauge-fixing and
  spacetime reconstruction in the hamiltonian theory of cosmological
  perturbations},\ }\href {https://doi.org/10.1088/1361-6382/aca385} {\bibfield
   {journal} {\bibinfo  {journal} {Classical and Quantum Gravity}\ }\textbf
  {\bibinfo {volume} {40}},\ \bibinfo {pages} {015003} (\bibinfo {year}
  {2022})}\BibitemShut {NoStop}%
\bibitem [{\citenamefont {{Sheth}}\ \emph {et~al.}(2001)\citenamefont
  {{Sheth}}, \citenamefont {{Mo}},\ and\ \citenamefont
  {{Tormen}}}]{ShethTormen}%
  \BibitemOpen
  \bibfield  {author} {\bibinfo {author} {\bibfnamefont {R.~K.}\ \bibnamefont
  {{Sheth}}}, \bibinfo {author} {\bibfnamefont {H.~J.}\ \bibnamefont {{Mo}}},\
  and\ \bibinfo {author} {\bibfnamefont {G.}~\bibnamefont {{Tormen}}},\
  }\bibfield  {title} {\bibinfo {title} {{Ellipsoidal collapse and an improved
  model for the number and spatial distribution of dark matter haloes}},\
  }\href {https://doi.org/10.1046/j.1365-8711.2001.04006.x} {\bibfield
  {journal} {\bibinfo  {journal} {\mnras}\ }\textbf {\bibinfo {volume} {323}},\
  \bibinfo {pages} {1} (\bibinfo {year} {2001})},\ \Eprint
  {https://arxiv.org/abs/astro-ph/9907024} {arXiv:astro-ph/9907024 [astro-ph]}
  \BibitemShut {NoStop}%
\bibitem [{\citenamefont {{Ostrowski}}\ and\ \citenamefont {{Delgado
  Gaspar}}(2022)}]{JOIDG}%
  \BibitemOpen
  \bibfield  {author} {\bibinfo {author} {\bibfnamefont {J.~J.}\ \bibnamefont
  {{Ostrowski}}}\ and\ \bibinfo {author} {\bibfnamefont {I.}~\bibnamefont
  {{Delgado Gaspar}}},\ }\bibfield  {title} {\bibinfo {title} {{On the maximum
  volume of collapsing structures}},\ }\href
  {https://doi.org/10.1088/1475-7516/2022/04/059} {\bibfield  {journal}
  {\bibinfo  {journal} {\jcap}\ }\textbf {\bibinfo {volume} {2022}},\ \bibinfo
  {eid} {059} (\bibinfo {year} {2022})},\ \Eprint
  {https://arxiv.org/abs/2112.05245} {arXiv:2112.05245 [gr-qc]} \BibitemShut
  {NoStop}%
\end{thebibliography}%
\end{document}